\documentclass[prl,twocolumn,amsmath,amssymb,longbibliography]{revtex4-1}
\usepackage{graphicx}
\usepackage{amsmath,amsfonts}
\usepackage{dcolumn}
\usepackage{bm}
\usepackage{slashed}
\usepackage{nicefrac}
\usepackage{epstopdf}
\usepackage{color}
\usepackage{tikz-feynman,contour}
\def\bra{\langle}
\def\ket{\rangle}

\def\cM{\mathcal{M}}

\def\cH{\mathcal{H}}
\def\cP{\mathcal{P}}
\def\cS{\mathcal{S}}

\def\vx{\mathbf{x}}

\def\C{\mathbb{C}}

\def\R{\mathbb{R}}
\def\Z{\mathbb{Z}}

\def\tr{\mathrm{Tr}}

\newtheorem{definition}{Definition}

\begin{document}
\title{Combinatorial Spacetime from Loop Quantum Gravity}
\author{Mikhail Altaisky} 
\affiliation{Space Research Institute RAS, Profsoyuznaya 84/32, Moscow, 117997, Russia}
\email{altaisky@cosmos.ru}
\date{}
\begin{abstract}
Loop quantum gravity is a perspective candidate for the  quantum theory of gravity. However, 
there is a conceptual controversy in it: having started from the Einstein-Hilbert action and 
describing spacetime without matter, we can hardly define spacetime as 
anything other than a set of relations between matter fields. Here, following the Penrose idea of combinatorial 
spacetime we reformulate loop quantum gravity theory solely in terms of the matter fields. 
\end{abstract}
\maketitle
\section{Introduction}
Unification of quantum mechanics (QM) and general relativity (GR) seems to be the most challenging problem of physics 
for more than a hundred years. Quantum mechanics, which arose from making the Poisson brackets of classical mechanics into the  
commutators of operators in QM, inherits the privileged  role of time $t$ as a crux of the theory. General relativity, in 
contrast, is a theory invariant under the general coordinate transformations $\mathrm{Diff}(4)$:
\begin{equation}
{x'}^{\mu}=f^\mu(x),  \label{gct}
\end{equation}
it should keep all the coordinates $\mu = 0,1,2,3$ on the same footing, with $x^0=ct$ being the time coordinate with 
no chance of getting any preferences. Since the problem was first understood, there is a pending question: What is more 
important GR or QM?

The known approaches to quantization of gravity can be loosely divided in two classes: 
(a) unified theories, where the gravity force emerges from some universal interaction on the same footing as the electroweak and strong interactions; (b) geometric 
approach, based on the $Diff(4)$ invariance of general relativity, where the gravity 
is considered separately from other physical interactions. The former approach have 
been inspired by superstring theory \cite{Green1984,GSW2012book}, which was later extended to $M$-theory \cite{Duff1996,BBS2007}, and is closely related to supersymmetry. The lack of experimental evidence of any particle superpartners 
is a real challenge to it. The latter approach is essentially based on general coordinate invariance \eqref{gct} of GR, and on the Einstein-Hilbert action functional, 
which stems from it:
\begin{equation} 
S_{EH} = \frac{1}{16\pi G} \int_\cM d^4 x \sqrt{-g} (R-2\Lambda) , \label{EH4}
\end{equation} 
where $R$ is  the Ricci scalar, and $\Lambda$ is the cosmological constant.

The most known, although not commonly accepted, approach of the second class 
is the {\em loop quantum gravity} (LQG) \cite{RS1995,Rovelli-book,Vidotto2014}. LQG does not pretend to describe the matter fields, but demonstrates interesting results in 
quantization of area and volume  \cite{RS1995,Rovelli-book,Vidotto2014}. 
The boundary between these two classes is rather dim, and crosses group field theory 
(GFT) and matrix models \cite{Freidel2005}, and also a number of other approaches, which we are not going to list here. The 
problem of a mutual consistence between LQG and string theory is also a matter of 
active discussion \cite{VS2024}.
An important feature of both approaches is the assumption of a continuous 
differentiable manifold. This may lead to contradictions in quantum physics, 
and still it is a sign of more credit to GR than to QM.

In GR the spacetime is described by metric tensor $g_{\mu\nu}(x)$, which can 
be determined, at least in principle, from the solution of Einstein equations 
\begin{equation}
R_{\mu\nu} - \frac{1}{2} g_{\mu\nu}R + \Lambda g_{\mu\nu} = \frac{8\pi G}{c^4} T_{\mu\nu}. \label{EE}
\end{equation}
where $T_{\mu\nu}$ is the energy-momentum tensor of the matter fields, and $G$ is the 
Newton constant. That is, if we know $T_{\mu\nu}$, the metrics and the curvature can 
be determined. 

In quantum theory, and specially in LQG, the quantization procedure starts with a 
''pure spacetime'' \eqref{EH4}, and it is not clear 
what is quantized, since there are no matter fields  ($T_{\mu\nu}\equiv0$), 
but any quantum measurement of length or time can happen only by virtue of matter fields. The problem was properly understood quite a long time ago, and the proposed 
solution was to express all metric properties solely in terms of 
measurable quantities \cite{Penrose1971}. This has become known as the Penrose's 
{\em spin networks} (SN) and the {\em combinatorial spacetime}, respectively. 

Despite a very elegant formal scheme for the construction of combinatorial spacetime,
its practical realization was given only in non-relativistic settings \cite{Penrose1971}. In this paper we provide a locally Lorentz-invariant 
realization of the Penrose combinatorial spacetime using relativistic spin networks of matter fields.
The novelty of the research is that instead of a continuous holonomy operator, 
usually implied by loop quantum gravity, we use a kind of Regge discretization 
with all the curvature effects concentrated at vertices of the interaction graph.  

\section{Spin networks and quantum geometry}
The concepts of the {\em combinatorial spacetime} and the {\em spin networks} both were introduced in \cite{Penrose1971}. The question was what is the meaning of {\em direction} and what is the meaning of {\em angle between two directions} in non-relativistic quantum mechanics. 

\begin{figure}[ht]
	\centering \includegraphics[width=4cm]{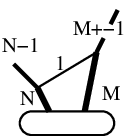}
	\caption{Transforming a qubit from $N$-block to $M$-block may result in either $(M\pm1)$-blocks. Redrawn from \cite{Penrose1971}}
	\label{mnd:pic}
\end{figure}

Suppose we have an electron, 
or any other spin-$\frac{\hbar}{2}$ particle. A measurement of its spin projection to an arbitrary chosen coordinate 
axis gives either of two possible values $\pm\frac{\hbar}{2}$. If we have a {\em pair of electrons}, the projection 
of the total spin may be either zero -- for the singlet state of the pair, or may be chosen with different 
probabilities from $(-\hbar,0,\hbar)$ for a triplet state. There are no directions associated with singlet state -- it is rotationally invariant -- but there is one direction associated with  triplet state. For a system of $N$ spin-$\frac{\hbar}{2}$ particles the maximal spin is $\frac{\hbar N}{2}$ and we have maximum $N+1$ possible projections of spin. 
Using addition rules for angular momenta, we can say something about the correlations between blocks of matter particles. If we manage to define a 'direction in space' associated with a block of a sufficiently large  number of spin-$\frac{\hbar}{2}$ we can also define the angle between  directions of 
two such blocks $M,N\gg1$. A scheme of such gedunken-experiment is drawn in Fig.~\ref{mnd:pic}.

The comparison of the directions of big blocks is 
defined operationally, as the interaction of  these blocks, such that we detach a single qubit from $N$-block and 
re-attach it to $M$-block, see Fig.~\ref{mnd:pic}.
We can expect two possible outcomes of such gedunken-experiment: either the $M$-block becomes $(M+1)$-block, or it becomes 
a $(M-1)$-block. In the former case we conclude that the blocks were {\em parallel} before the experiment has been accomplished, in the latter -- that 
they were anti-parallel. Here $N$-block is understood as an entity capable of having $N+1$ projections of its spin. In general, the probability of getting the $(M+1)$ block defines the angle $\theta$ between the 
directions of spins of two blocks 
\begin{equation}
P(\theta) = \frac{1}{2}(1+\cos\theta).
\end{equation}
Having sufficient number of copies of such blocks, we can estimate the angles between blocks as the rational probabilities $P=\frac{m}{n}$ of getting $(M+1)$-blocks from $M$-ones; $m$ is the number of successes, $n$ is the total number of experiments. 
So the directions can be introduced only if we have some blocks of matter particles 
with sufficiently high total spin. There are no 'directions' for a single spin-
$\frac{\hbar}{2}$ fermion. 

The scheme drawn in Fig.~\ref{mnd:pic} does not include time -- it includes only {\em objects} and transformations of these objects. In the modern language of {\em category theory} the the edges of the graph Fig.~\ref{mnd:pic} -- blocks of spins -- are the {\em objects}, and the vertices of the graph are {\em morphisms} \cite{Beer2018}. The absence of time is quite natural, since the method is based 
on the angular momentum operator, defined by only one fundamental constant $\hbar$:
\begin{equation}
[ \hat{L}_i,\hat{L}_j]=\imath \hbar \epsilon_{ijk}\hat{L}_k.
\end{equation}
To measure the time and length the other constant $c$ is required.

Each spin-$\frac{\hbar}{2}$ particle [or a qubit] in such diagrams is represented by solid line. A $N$-qubit block is represented by a strand of $N$ such lines. 
The diagrams denoting the transition between such blocks are known as Penroses's spin networks. To provide the angular momentum 
conservation in each vertex the fermion line entering a vertex from one strand should exit from another; reversal 
paths through the same strands are not allowed.  
\begin{figure}[ht]
	\centering \includegraphics[width=4cm]{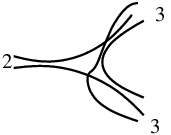}
	\caption{Vertex structure in a spin network}
	\label{sn:pic}
\end{figure}

Original spin networks were the quantum schemes of operations performed on 
blocks of spin-$\frac{\hbar}{2}$ particles, such that at each transformation vertex 
the angular momentum is conserved. In other words it is a graphical technique describing manipulations on blocks of particles, transforming under rotations 
according to representations of $SU(2)$ group, so that each interaction 
vertex [intertwiner] multiplied by all connected edges is invariant under $SU(2)$ rotations. Later it was generalized to arbitrary Lie group. 
Formally, a spin network 
\begin{equation}
\cS = (\Gamma, j_l, n_r)
\end{equation}
is a graph $\Gamma$, with a set of edges $\{ j_l \}_l$, labelled by representations of a Lie group $G$, and a 
set of vertices $\{ n_r \}_r$, also called the intertwiners, constructed so, that combined with all incoming edges they form
singlet  of $G$. To each edge $j_l$, connecting $n_i$ and $n_j$ vertices, we can associate a group element $g_{ij}\in G$, 
so that $g_{ij}=g_{ji}^{-1}$. 

Spin network technique, based on the representations of $SU(2)$ group, 
has entered gravity theory in a way very different from the original 
Penrose's proposal. Namely, from an abstract mathematical idea of 3d gravity, a functional invariant under $\mathrm{Diff}(3)$ transformations 
\begin{equation}
S[g] = \int_{\cM_3} d^3x \sqrt{g} R, \label{EH3}
\end{equation}
it was observed by Ponzano and Regge \cite{Ponzano1968}, 
that the Einstein-Hilbert action \eqref{EH3} can be approximated 
by a discrete sum 
\begin{equation}
S_{\mathrm{Ponzano-Regge}} = \sum_\mathrm{tetrahedra}\sum_{i=1}^6 \theta_i \left(j_i + \frac{1}{2} \right), \label{rpt}
\end{equation}
over triangulations of a manifold $\cM_3$ into tetrahedra 
with the edge length [taken in appropriately small units] $l_i = j_i +\frac{1}{2}$, with  $\theta_i$ being the angle between the outward normals to  two faces joined by the $i$-th edge. The half-integers $j_i$ were identified with the spin representations 
of $SU(2)$ rotation group by means of the Racah coefficients \cite{Ponzano1968}. A single tetrahedron of the tessellation 
\eqref{rpt} with edges labelled by 'spin indices' $j_k$ is schematically shown in Fig.~\ref{t3:pic}.
 \begin{figure}[ht]
 	\centering \includegraphics[width=4cm]{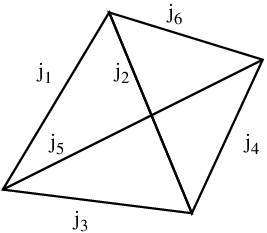}
 	\caption{Tetrahedron labelled by spin indices on its edges}
 	\label{t3:pic}
 \end{figure}
There is nothing about the angular momentum in \eqref{EH3} by construction -- it is only about general coordinate transformations -- 
and the 'spin indices'  $j_k$ have the meaning of length. There is also 
nothing about any particles, in contrast to that originally proposed by Roger Penrose \cite{Penrose1971}.

A breakthrough in the application of spin network technique to gravity has been achieved when the metric argument $g_{\mu\nu}(x)$ of the Einstein-Hilbert action \eqref{EH4} have been changed to the 
Ashtekar connection $A_\mu^{IJ}$ and the tetrads $e_\mu^I$ \cite{Ashtekar1986}.

All our local physical measurements are described by quantum field 
theory amplitudes calculated in a flat spacetime with Minkowski  
metrics $\eta_{IJ}=\mathrm{diag}(1,1,1,-1)$. To describe a curved 
4d spacetime $\cM_4$ we can use one-forms 
$$
e^I(x) = e^I_\mu(x) dx^\mu.
$$
The matrices $e^I_\mu$ are known as {\em tetrads}. If $\cM_4$ is flat 
and coincides with Minkowski space $e^I_\mu\equiv \delta^I_\mu$, otherwise $e^I_\mu$ is a function of $x \in \cM_4$. 
The usual metrics $g_{\mu\nu}(x)$ is expressed in terms of tetrads as 
\begin{equation}
g_{\mu\nu}(x) = e^I_\mu(x) e^J_\nu(x) \eta_{IJ}. \label{tet}
\end{equation}

Parallel transport of vector $e^I$ from a point $x$ to $x+dx$ on $\cM_4$ is given 
by covariant derivative 
$$ D e^I = d e^I + \omega^I_{\phantom{I} J}\wedge e^J,$$
where $ \omega^I_{\phantom{I} J} = \omega^I_{\mu J} dx^\mu $ is a 
spin-connection taking values in $so(3,1)$ Lie algebra. 

If $\omega$ 
is the $SO(3,1)$ spin connection, defined to keep the tetrads covariantly constant 
\begin{equation}
\partial_{[\mu} e^I_{\nu]} + \omega^I_{[\mu J}[e] e^J_{\nu]}=0,
\end{equation}
the action \eqref{EH4} takes the form
\begin{equation}
S[e,\omega] = \frac{1}{16\pi G} \int d^4 x e e^\mu_I e^\nu_J R^{IJ}_{\mu\nu}[\omega],
\end{equation}
where 
$$
R^\mu_{\phantom{\mu}\nu\tau\sigma}[g[e]] = e^\mu_I e_{\nu J} R^{IJ}_{\tau\sigma}[\omega[e]]
$$
is curvature of the connection $A$, 
and $e \equiv \mathrm{det}(e^\mu_I)$.
Constructing the complex-valued Ashtekar connection
\begin{equation}
A^{IJ}_\mu [\omega] = \omega^{IJ}_\mu - \frac{\imath}{2} \epsilon^{IJ}_{\phantom{IJ}MN} \omega^{MN}_\mu,
\end{equation}
we can rewrite Einstein-Hilbert action in a form of gauge theory 
\begin{equation}
S[e,A] = \frac{1}{16\pi G} \int d^4x e_{\mu I} e_{\nu J} F^{IJ}_{\tau\sigma}[A]\epsilon^{\mu\nu\tau\sigma},
\end{equation}
where
$$
F^{IJ}_{\mu\nu}[A] = R^{IJ}_{\mu\nu}[\omega] - \frac{\imath}{2} \epsilon^{IJ}_{\phantom{IJ}MN} R^{MN}_{\mu\nu}[\omega],
$$
see, e.g., \cite{Rovelli1991CQG} for details.

Being written in  the Ashtekar variables, the quantum state of spacetime turns to be the state of $SL(2,\C)$ connection $A$, 
described by a vector $\psi(A)$ in Hilbert space of states. For this reason, the quantization in the Ashtekar variables is also known as {\em connectodynamics} -- in contrast to {\em geometrodynamics} of 
$(g_{ab},\pi^{ab})$-quantization.

 The connection $A$ is naturally  
characterised by  holonomy operator 
\begin{equation}
U_\gamma [A] = \cP e^{\imath \int_\gamma ds \dot{\gamma}^\mu (s) A_\mu^B(x(s)) T_B}, \label{hol1}
\end{equation}
where the integration is performed over a curve $\gamma = \gamma(s)$, $\dot{\gamma}^\mu(s)$ is a tangent vector along the curve $\gamma(s)$, with $\cP$ being the path-ordering operator, and $T_B$ being the generators of the symmetry group in the representation appropriate 
to connection $A$. [This may be $SO(3)$ rotation group for 3d Regge gravity, $SL(2,\C)$ group for the spin connection, etc.]  
Each state of geometry, i.e. how the given connection $A$ rotates the matter fields, can be written in a loop 
basis $\{ |\alpha\ket \}$:
\begin{equation}
\psi_\alpha[A] = \tr (U_\alpha[A]), \label{wf1}
\end{equation}  
where $\alpha$ runs over a set of all possible loops. The basis of loops $\{ |\alpha\ket \}$ is overcomplete and forms 
a frame \cite{RS1995}. The use of the loop basis to study a theory invariant under diffeomorphisms is quite common in differential geometry \cite{Dubrovin1985}. Loop quantum gravity, additionally to this common ground, has deeper roots in graph 
theory \cite{CKY1997}. 

\begin{figure}
	\centering \includegraphics[width=4cm]{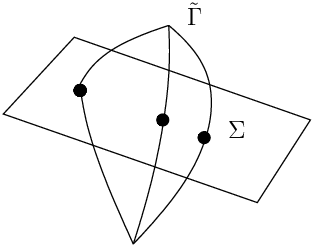}
	\caption{Spin network graph $\tilde{\Gamma}$ crossing the surface $\Sigma$ in $n=3$ points}
	\label{sigma:pic}
\end{figure}

The quantization of the theory based on the Ashtekar connection 
naturally leads to the quantization of area, which is simplest in 
$d\!=\!3$ dimensions. If we have a two-dimensional surface $\Sigma$, 
the area of this surface is given by the area vector 
\begin{equation}
L^I_\Sigma = \frac{1}{2} \epsilon^I_{\phantom{I}JK} \int_\Sigma e^J \wedge e^K, \label{AO}
\end{equation}
where $e^J = e^J_\mu dx^\mu$ is  one-form. 
If surface $\Sigma$, is spanned by two vectors $e_x$ and $e_y$ 
tangent to this surface, the area of $\Sigma$ is given by the integral of absolute value of  vector product $e_z = e_x \times e_y$ over this surface, or, in a component notation 
\begin{equation}
S = \int_\Sigma d^2x \sqrt{\delta_{IJ} e^I_z e^J_z}. \label{s2c}
\end{equation}
In canonical quantization of gravity, the Ashtekar variables $(A^a_I,e_b^J)$ are considered as  canonical coordinates and their conjugated momenta, respectively. This means, in quantum theory we are to make the substitution 
$$
e^I_a \to -\imath \hbar \frac{\delta}{\delta A^a_I}.
$$
The variation of the operator 
(\ref{hol1}) with respect to canonical coordinate gives
\begin{equation}
\frac{\delta}{\delta A^a_I} U_\gamma[A] = \imath e_a T^I U_\gamma[A].
\end{equation}
This means that in quantum theory, the counterpart of a classical 
equation \eqref{s2c} takes the form 
\begin{equation}
\hat{A}_\Sigma \Psi_{\tilde{\Gamma}} = \sum_{p=P_1}^{P_n} 
\sqrt{- \delta_{IJ} \frac{\delta}{\delta A^z_I} \frac{\delta}{\delta A^z_J}} \Psi_{\tilde{\Gamma}}, \label{A2q}
\end{equation}
where $\hat{A}_\Sigma$ is the area operator of the surface $\Sigma$, 
$\Psi_{\tilde{\Gamma}}$ is the wave function of quantum geometry, 
determined by connection $A$ for a spin network with graph $\tilde{\Gamma}$, 
intersecting the surface $\Sigma$ in $n$ points, as shown in 
Fig.~\ref{sigma:pic},
Wave function of such geometry, determined by the connection $A$, can be written in the form 
\begin{equation}
\Psi_{\tilde{\Gamma}} = \Psi(U_1,U_2,\ldots,U_N), \label{gwfN}
\end{equation}
where $U_k[A]$ denotes the holonomy operator for 
the $k$-th edge of graph $\tilde{\Gamma}$, which crosses the 
surface $\Sigma$. The scalar product of two geometry states on a 
graph $\tilde\Gamma$ is defined in a usual way \cite{Rovelli-book}:
$$ \bra \Psi_{\tilde\Gamma}| \Phi_{\tilde\Gamma}\ket = \int dU_1\ldots dU_N 
\bar{\Psi}(U_1,\ldots,U_N) \Phi(U_1,\ldots,U_N).$$

In case of the $SO(3)$ group, the contraction of Roman indices $I,J$ in 
\eqref{A2q} 
gives the squared angular momentum operator $\hat{J}^2 = \sum_{I=1}^3 (T^I)^2$, from where 
it follows that area operator  $\hat{A}_\Sigma$ takes a discrete set of eigenvalues: 
\begin{equation}
\hat{A}_\Sigma \Psi_{\tilde{\Gamma}} = \sum_{p=P_1}^{P_n} \sqrt{\hat{J}^2} \Psi_{\tilde{\Gamma}} = \sum_{p=P_1}^{P_n} \sqrt{j_p(j_p+1)} \Psi_{\tilde{\Gamma}}.
\end{equation}
Similar, but more involved  consideration is valid for the volume 
operator \cite{RS1995NPB,AL1997}. The problem here is that the surface 
$\Sigma$ is not yet defined in terms of the spin network $\tilde{\Gamma}$.

Quantization of spacetime, or as is commonly accepted quantization 
of a continuous differentiable manifold endowed with Einstein-Hilbert action, stems from the Regge observation, that a tessellation 
of compact 3d manifold by 3d simplices -- tetrahedra -- is  related to $SO(3)$ rotation 
group by means of Racah coefficients \cite{Ponzano1968}. A two-dimensional sphere $S^2$ can be partitioned into 2d simplices -- 
triangles, 3d compact manifold -- into tetrahedra, 4d manifold -- into 4d simplices, and so on. Providing these 
simplices be small enough, all the curvature information about the manifold is encoded in deficiency angles of the 
tesselation: at vertices in 2d, at edges in 3d, at faces in 4d. The edges and faces are considered 
to be flat by this triangulation procedure \cite{Regge1961}. A graph $\tilde{\Gamma}$ connecting the centres of 
simplices in the tessellation is considered as a spin network (SN) graph; the simplices -- as the quanta of space. 
In case of a three-dimensional manifold, the tetrahedra are glued to each other by joining  their faces -- triangles, and 
the SN is a thread network, which keeps the tetrahedra together.
A fragment of triangulation of a compact 3d manifold by tetrahedra is shown in Fig.~\ref{fig1:pic} below.
\begin{figure}[ht]
	\centering \includegraphics[width=6cm]{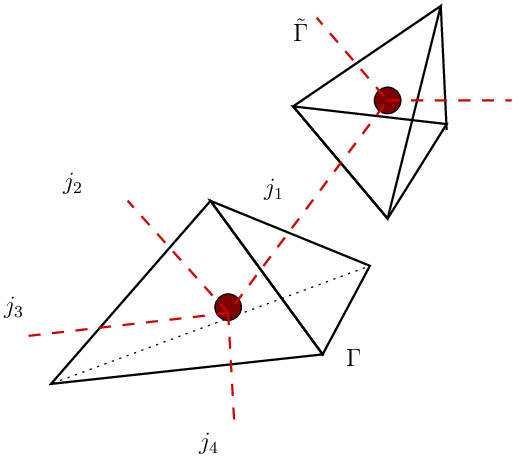}
	\caption{Partitioning of a 3d  manifold into tetrahedra. Triangulation graph $\Gamma$, consisting 
		of the edges of tetrahedra and their vertices, is shown by black solid lines. The dual graph $\tilde{\Gamma}$, which 
		connects the centres of the tetrahedra, is shown by  dashed lines.}
	\label{fig1:pic}
\end{figure}
The elementary area is that of a triangle crossed by the spin network 
graph $\tilde{\Gamma}$.

The physical problem here is that if the spacetime {\em is not a continuous differentiable manifold}, we will not be able to construct 
both the triangulation graph $\Gamma$ and the spin network graph $\tilde{\Gamma}$ simultaneously, but will have to consider only one 
graph.

\section{Matter spin networks}
Let us recall the definition of manifold \cite{Dubrovin1985}:
\begin{definition}
	Differentiable $n$-dimensional manifold is an arbitrary set of points $M$, endowed with the following structure:
	\begin{enumerate}
		\item $M$ is a union of a finite or countable set of sets $U_q$, 
		referred to as charts.
		\item $\forall U_q$ there is a set of coordinates  $x_q^\alpha, \alpha = \overline{1,n}$, referred to as local maps. The (non-zero) intersection of two charts $U_p \cap U_q$ is a chart itself, where two maps are mutually consistent:
		$$
		x^\alpha_p = 	x^\alpha_p (x_q^1,\ldots,x_q^n), \quad 
	    x^\alpha_q = 	x^\alpha_p (x_p^1,\ldots,x_p^n),
	    $$ so that 
		$ \mathrm{det} \bigl(
			\frac{\partial x^\alpha_p}{\partial x^\beta_q} \bigr)\ne 0
		$.
	\end{enumerate}
\end{definition}

From physical point of view, the coordinates can be assigned only to 
{\em events}, the changes of physical particles or systems that can 
be observed. There may be no coordinates without events \cite{MTW}.
This means, in a physical theory of spacetime, the charts $U_q$ should be the subsets of the set of all events in the Universe.

Considering spacetime as a set of interaction vertices, labelled by 
coordinates, is in agreement with the Penrose notion of combinatorial 
spacetime. Taking the matter particles as {\em objects} [of the category theory] and the interaction vertices as {\em morphisms} we arrive at a 
usual Feynman diagram describing the world history. 
In other words, we have a triangulation graph $\Gamma$, the edges of which are labelled by matter particles. There is nothing here about 
the dual graph $\tilde{\Gamma}$. That is why it is desirable to reformulate LQG in terms of the graph $\Gamma$, with  physical particles associated to its edges . 

In a usual quantum field theory the edges of Feynman diagrams are identified with representations of the Poincar\'e group multiplied by 
approapriate representation of the internal symmetry group 
\cite{Weinberg1}. We cannot do it in a curved space, since the 
Poincar\'e group assumes the invariance under spacetime translations -- 
that is a {\em flat} space. Instead, we need something more general 
which will reduce to the Poincar\'e group in the limit of flat space.

Similarly to usual LQG, a complicated triangulation graph $\Gamma$ can 
be composed of oriented elementary loops, which we associate with elementary fermions of spin $\frac{1}{2}$. We can identify fermions with $SL(2,\C)$ spinors, periodic in the loop 
parameter: $$u(s)=u(s+2\pi).$$

There will be a difference from usual SN at this point. In the Penrose SN technique only the total angular momentum of a graph edge matters, 
providing the angular momentum conservation in each vertex, with 
all braiding information attributed to the vertices. Here we care of 
the spin states of particles on the edges of graph, and denote the spin $J=1$ states by two parallel lines ($=$), and the singlet state $J=0$  by two crossed lines   ($\times$). Both states are present in the product of two spin-$\frac{1}{2}$ fermions 
$$D_\frac{1}{2}\otimes D_\frac{1}{2} = D_1 \oplus D_0.$$

\begin{figure}[ht]
 	\centering \includegraphics[width=6cm]{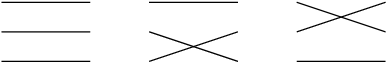}
 	\caption{Decomposition of a product of three fermion lines 
 		$D_\frac{1}{2} \otimes D_\frac{1}{2} \otimes D_\frac{1}{2}$ into a sum of representations. 
 		There are no arrows on the lines since the directions of loops are not yet fixed.}
 	\label{d32:pic}
 \end{figure}

 Similarly, for three fermion lines we have 
 the equation 
 \begin{equation}
 D_\frac{1}{2}^3 = D_\frac{3}{2} \oplus 2 D_\frac{1}{2},
 \end{equation} 
 the r.h.s. of which corresponds to the diagrams shown in Fig.~\ref{d32:pic}. 

Our goal is to construct four-dimensional spacetime from elementary matter fields, fermionic loops. For a single loop without any interaction vertices 
nothing can be measured. Non-interacting loop is merely a mapping 
from $S^1$ to the Hilbert space of states of matter fields: 
 \begin{equation}
 S^1 \stackrel{u}{\to} \cH, \quad u(s) = u(s+2\pi). \label{lm}
 \end{equation}
  So, non-interacting loop can be described as  
 \begin{equation}
 u(s) = \begin{pmatrix} u_1(s) \cr \vdots \cr u_N(s) \end{pmatrix}, 0 < s \le 2\pi \label{mcwf}
 \end{equation}
 where $N$ is dimension of the representation.
 Unless there are vertices [{\em events}] on the loop, parameter $s$ is not observable, and 
 all physical observables should be invariant under translations of loop parameter. 
 Functions on loops therefore can be decomposed in Fourier series, i.e., with respect to representations of the translation group 
 on the loop, 
 \begin{equation}
 u(s) = \sum_m C_m e^{\imath m s}, \quad m \in \Z, \label{pr}
 \end{equation}
 where the ''momentum'' $m$ coincides with the Noether current corresponding to the 
 $U(1)$ phase transformation 
 $$u \to e^{\imath\alpha} u, \alpha \in \R$$
 of a free-field action.
 
 Loop parameter $s$ in our case cannot be directly understood as  ''length'' -- as it happens for usual Fourier transform in quantum field theory. Since $s$ is not a measurable quantity, it should be understood as an artefact of periodicity of  the loop  mapping \eqref{lm}. Any definition of physical distance [or interval] should rely on events -- vertices 
 of the graph $\Gamma$.
 
\begin{figure}[ht]
 	\centering \includegraphics[width=6cm]{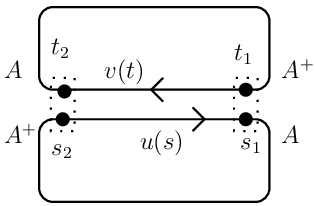}
 	\caption{Constructing a spin-1 boson from two identical fermion loops with spin $\frac{1}{2}$. 
 		The two vertices are identified with the value of loop parameters $(s_1,t_1)$ and $(s_2,t_2)$, respectively. 
 		The connection matrix of the first loop, at the vertex $s_1$ is $A$. It results in the rotation of 
 		the field $u(s)$ at $s_1$ according to: $u(s_1+\epsilon)=A u(s_1-\epsilon)$}
 	\label{l2:pic}
 \end{figure}

 If we have two identical loops of spin $\frac{1}{2}$ we can  construct 
 a spin-1 boson by joining these two loops in two interaction vertices, 
 as shown in Fig.~\ref{l2:pic}. The representation of a spin 1 
 boson as a tensor product of two spinors $p_{\alpha\dot{\alpha}} = \lambda_\alpha \tilde{\lambda}_{\dot{\alpha}}$ is quite common 
 in quantum field theory, see e.g. \cite{HHH2021}. 
By definition, the interaction vertices  are operators (morphisms) that change the states 
of  incoming matter fields into the states of outgoing fields. There are two identical vertices in the diagram Fig.~\ref{l2:pic} (shown by dotted lines), labelled by loop parameters $(t_1,s_1)$ and $(t_2,s_2)$, respectively. Let us change the state of the $u(s)$ at vertex 1 at 
the values of loop parameter $s=s_1$.  This means 
\begin{equation}
u(s_1+\epsilon) = A u(s_1-\epsilon), \epsilon \to +0, 
\end{equation} 
where $A$ is the transformation matrix for the field $u$. The periodic 
property of the loop \eqref{lm} should be observed at the presence 
of interaction vertices as well. This means the transformation matrix $A^+$ at 
the second vertex, applied at the value of parameter $s=s_2$, should 
return the field $u(s)$ to its original state:
$$
u(s_2+\epsilon)= A^+ u(s_1+\epsilon)=A^+ A u(s_1-\epsilon),
$$
from where it follows that $A^+ A=\mathbb{I}$, since the changes take 
place only in vertices, and $u(s_2+\epsilon)=u(s_1-\epsilon)$. The 
same consideration is true for the other loop parametrised by $t$.

Considering a limiting case of usual QED, with the counterpart of diagram Fig.~\ref{l2:pic} being the vacuum diagram,
$$
\begin{tikzpicture}[baseline=(e)]
\begin{feynman}[horizontal = (a) to (e)]
\vertex  (a) at (0,0);
\vertex [right =2cm of a] (e);
\diagram* {
	(a) -- [half left, fermion] (e) -- [half left, fermion] (a),
	(a) -- [photon] (e), 
};
\end{feynman}
\end{tikzpicture}
$$
 we can take $u$ and 
$v$ as four-component Dirac spinors $u = \begin{pmatrix}
u_L \cr u_R
\end{pmatrix}$, so that the transformation matrix $A$ takes the form 
$$
A = \begin{pmatrix}
e^{\frac{\imath}{2}(\vec{\omega}-\imath \vec{\nu})\vec{\sigma}} & 0 \cr
0 & 
e^{\frac{\imath}{2}(\vec{\omega}+\imath \vec{\nu})\vec{\sigma}}
\end{pmatrix},
$$
and its inverse $A^+$ is related to the Hermitian conjugated matrix 
by parity transformation 
$$
A^+ = \begin{pmatrix}
0 & 1 \cr 1 & 0
\end{pmatrix}  A^\dagger 
\begin{pmatrix}
0 & 1 \cr 1 & 0
\end{pmatrix}.
$$

Considering spinors on a loop we have neither introduced the 'length' 
or the 'mass' parameter. Both are related to the group of translations.
In usual QFT, particles on the edges of Feynman diagrams are 
identified with representations of Poincar\'e group. The mass parameter 
$M$ labels these representations, since $\hat{p}^2=M^2$ is the Casimir 
operator of the Poincar\'e group, with $\hat{p}$ being the generator of 
4d translations. In case of loops we do not have 4d translations, but 
only 1d translations along the loop: $$t \to t + a\ \mathrm{mod\ }2\pi.$$

An elementary fermion loop without any interaction vertices cannot be 
subjected to any observation, so its length can be set to some {\em minimal distinguishable length} $l_0$ by definition; presumably a Planck 
length $l_0 \propto l_{Pl}$. Considering the interaction of two loops in two 
vertices, as shown in Fig.~\ref{l2:pic}, we can distinguish two 
different parts in each loop, and it is natural to assume that all of 
them have the same minimal length $l_0$. Thus the length of each loop in 
Fig.~\ref{l2:pic} is $2l_0$ due to the introduction of two vertices, 
changing their states. The same is true for arbitrary number of vertices: if we endow a loop with $N$ vertices its length becomes $N l_0$.

It is natural to measure the lengths in units of action $\hbar$. 
Momentum operator $\hat{p} = -\imath \hbar \frac{\partial}{\partial x}$ in quantum mechanics is the generator of translations:
$$
\psi(x+\delta x) = \psi(x) + \delta x \frac{\partial \psi}{\partial x} 
+ \ldots = \left(1+ \imath \frac{p\delta x}{\hbar}\right)\psi(x) 
+ \ldots$$
So, the change of the wave function $\psi(x)$ cased by the translation on $\delta x$ is proportional to the integer number of minimal quanta 
($\hbar$) in the action $p \delta x$ required for this translation. 
If a loop is endowed with $N$ interaction vertices its length is 
$\lambda = N l_0$ and the quantization condition along the loop 
is naturally the Bohr-Sommerfeld 
quantization condition 
\begin{equation}
\oint pdx =  2\pi\hbar N,
\end{equation} 
which determines the length of the loop. 
If we know its length $\lambda = N l_0$, its ''Compton'' mass  
will be 
\begin{equation}
M = \frac{h}{\lambda c}.
\end{equation}

Measuring lengths, or more exactly the {\em intervals}, in the units 
of action is in a fair agreement with classical limit. 
In relativistic mechanics, the action of a particle of rest mass $M$ moving along spacetime curve $(t=t(\tau),\vx = \vx(\tau))$ from $a$ to $b$, is proportional to the 
interval, or ''four-dimensional length'' of the curve: 
\begin{equation}  
S = -Mc \int _a^b ds = -Mc \int_a^b \sqrt{c^2 dt^2 - d\vx^2 }. \label{dS}
\end{equation}
For a particle of mass $M$, the substitution  
\begin{align}\nonumber 
p^0 = \frac{Mc}{\sqrt{1-\frac{v^2}{c^2}}}&,& p = \frac{Mv}{\sqrt{1-\frac{v^2}{c^2}}},\\
\nonumber dx^0 = cdt &,& dx =v dt
\end{align}
suggests that equation \eqref{dS} can be written in the form 
\begin{equation}
S = -\int_a^b \sqrt{\left(p_0 dx^0 - p  dx \right)^2} \equiv -\int_a^b \sqrt{\left(\sum_{\mu} p_\mu dx^\mu \right)^2}. \label{s01}
\end{equation}
Actions of two identical massive particles moving in 
opposite directions, comprising a 'photon' in Fig.~\ref{l2:pic}, 
identically cancel each other:
$$\int_a^b + \int_b^a = 0.$$
This keeps the {\em interval} of the 'photon' identically zero -- 
regardless particular value of $M$.

\begin{figure}[ht]
	\centering \includegraphics[width=8cm]{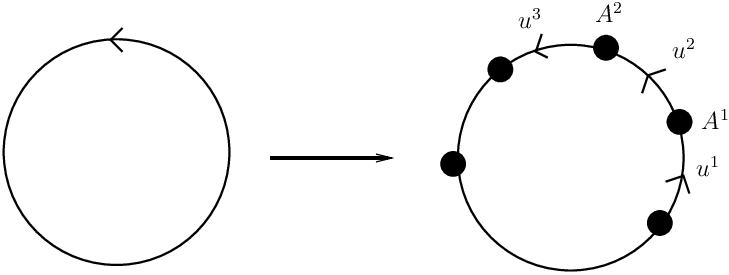}
	\caption{Adding a set of vertices to a free loop makes continuous Langrangian into a  
		discrete one}
	\label{mvl:pic}
\end{figure}

In a continuous theory of a field $u(s)$ on a loop, taken in 
certain $n$-dimensional representation of internal symmetry group, 
the translations of field along the loop would be given by covariant 
derivative 
\begin{equation}
\nonumber D_s = \partial_s + \mathcal{A}(s), 
\end{equation}
where $ \mathcal{A}(s)$ is $n\times n$ connection matrix, 
which generates field rotations in the internal space. 
Physically, at the absence of interaction vertices nothing 
can be observed on a free loop, and we ought to consider discretization 
where all changes of matter fields take place in a finite set of $N$ vertices:
$$
u(s_v+\epsilon) = \hat{A}(s_v) u (s_v-\epsilon), \epsilon\to +0,
s_v \in \{ s_1,\ldots s_N \}.
$$
In these settings, the continuous loop parameter $s$ does not have 
sense any longer, and we can write the evolution of fields along 
the loop in discrete form
\begin{equation}
u^{k+1} = \hat{A}^k u^k, \label{AA}
\end{equation}
as it is illustrated in Fig.~\ref{mvl:pic}. This a typical category 
where the matter fields $u^k$ are the objects, and the connections 
$\hat{A}^k$ are the morphisms. 
Let us consider a discretization of a  matter Langrangian 
for Grassmanian fields on a loop
\begin{equation}
L = \frac{\alpha_{ij}}{2} \left( \bar{u}_i \dot{u}_j - \dot{\bar{u}}_i u_j \right) + \imath \mu_{ij} \bar{u}_i u_j,
\end{equation}
where $\alpha_{ij}$ and $\mu_{ij}$ are Hermitian matrices, $i,j=\overline{1,n}$,
dot denotes derivative with respect to the loop parameter, and bar means complex conjugation.  For simplicity let us set $\alpha_{ij}=\delta_{ij}$,$\mu_{ij}=M \delta_{ij}$. In discrete case,  
the total change of the fields is given by \eqref{AA}, so the dot 
derivative becomes a finite difference:
\begin{eqnarray}\nonumber 
L = \frac{1}{2} \left( \bar{u}^k (u^{k+1}-u^k)
- (\bar{u}^{k+1}-\bar{u}^k)u^k
\right) + \imath M \bar{u}^k u^k, \\
\label{lk}
\end{eqnarray}
where the summation over internal indices is implied and the 
vertex number $k$ plays a role of discrete coordinate. 
Considering all matter fields $u^k$ and $u^{k+1}$ as independent, 
we can avoid using the connections $\hat{A}^k$, and write the 
Lagrangian  \eqref{lk} in a form 
\begin{equation}
L = \frac{1}{2} \left(\bar{u}^k u^{k+1} - \bar{u}^{k+1}u^k\right)
+ \imath M \bar{u}^k u^k.
\end{equation}
Consequently, if we integrate over all possible field configurations 
$(u^1,\ldots, u^N)$ we automatically integrate over all possible 
geometries $(\hat{A}^1,\ldots, \hat{A}^N)$ by virtue of \eqref{AA}.
To have the full action of the spacetime we have to sum up over all 
interacting loops with all their vertices. 

Consider a simple case of $n=2$, when the fields $u$ are $SL(2,\C)$ spinors and $\hat{A}^k$ are $SL(2,\C)$ transformations. The proper 
orthochronous Lorentz group $SO(3,1)$ is isomorphic to the group 
$SL(2,\C)/\Z_2$. Thus we can take either left or right spinors to represent Lorentz rotations. Let it be left spinors for definiteness.
Connection matrix $\hat{A}$ in any vertex can be decomposed in the basis of 
Pauli matrices:
\begin{equation}
\hat{A} = a_0 \mathbb{I} + \sum_{i=1}^3 a_i \hat{\sigma}^i, \quad 
\mathrm{det} A = a_0^2 - \sum_{k=1}^3 a_k^2=1. \label{pma}
\end{equation}
We can associate a tetrad vector $\mathbf{e}^0$ with the first fermion  
$u^1 = \begin{pmatrix}
u^1_1 \cr u^1_2
\end{pmatrix}$. Let the coordinates of this vector be arbitrary set to 
$\mathbf{e}^0=(1,0,0,0)$. Then we apply the connection matrix to 
the first fermion $u^1$ to get the state of the next particle 
$u^2 = \hat{A}u^1$. According to the correspondence between the spinor rotations 
and vector Lorentz rotations, the tetrad vector $\mathbf{e}^1$, corresponding to fermion $u^2$, will be given by 
\begin{equation}
e_\rho^1 = e_\mu^0 \Lambda^\mu_{\phantom{\mu}\rho}. \label{lte}
\end{equation}
The components of the $SO(3,1)$ Lorentz rotation matrix are given by \cite{Ruhl1967}:
\begin{align} \nonumber 
\Lambda^0_{\phantom{0}0} &=|a_0|^2+\sum_{k=1}^3 |a_k|^2,\\
\nonumber 
\Lambda^k_{\phantom{k}0} &= a_0 \bar{a}_k + \bar{a}_0a_k  + \imath \epsilon^{klm} a_l \bar{a}_m,\\
\Lambda^0_{\phantom{0}k} &= a_0 \bar{a}_k + \bar{a}_0a_k  - \imath \epsilon^{klm} a_l \bar{a}_m, \\
\nonumber 
\Lambda^l_{\phantom{l}k} &= \delta^l_k \left(|a_0|^2 - \sum_{k=1}^3 |a_k|^2\right) + a_k \bar{a}_l + \bar{a}_k a_l \\
\nonumber 
&+ \imath \epsilon^{klm} (\bar{a}_0 a_m - a_0 \bar{a}_m ). 
\end{align}
In this way, each edge of the triangulation graph can be associated with 
both the matter field $u^k$ [in accordance to the total spin 
of the given edge], and the tetrad vector $\mathbf{e}^k$. The coordinates of the 
latter should be chosen to get the same tetrad vectors when going along 
different loops. Having defined tetrad vectors ($\mathbf{e}^k$) in terms 
of the matter fields ($u^k$) we completely define geometry. 

Having defined tetrad vectors corresponding to the edges of triangulation graph $\Gamma$, we can also express the elementary areas as {\em  bivectors} \cite{BC1998}, composed of pairs of basic vectors following each other in the loop:
\begin{equation}
F^{[ij]} = \frac{1}{2}  \mathbf{e}^{[i,j]} \wedge \mathbf{e}^{[i+1,j]} , 
\hbox{with\ } \mathbf{e}^{[i+1,j]}_r = \mathbf{e}^{[i,j]}_m \Lambda^m_{\phantom{m}r},
\end{equation}  
where $j$ labels the loop, and $i$ labels the edge. The edges $i$ and $i+1$ are assumed to be adjusted by transformation matrix  $\hat{A}_j^i$ along the loop $j$. 

Spacetime in this picture is a function of quantum states of all matter in the universe, and in this sense there is no special force such as 'gravitation'. Perhaps, this does not rule out the existence of 
four-fermion line constructions in the network of interacting loops 
which can be understood as spin 2 'gravitons'. 

\section{A toy-universe model}
We do not know the topology of the Universe. The only things we 
know is that it is locally Lorenzian and is almost flat at the experimentally accessible energies. So, we cannot construct a realistic 
model of Universe triangulating a known manifold by matter particles,
like a triangulation of a compact manifold in a 3d Regge gravity \cite{Regge1961}. What we can do, is to construct toy models of a 
few fermion loops, combining them in a number of vertices.

Consider a toy model of a tetrahedron consisting of four fermion and 
two boson lines on the edges \cite{Altaisky24psn}, which corresponds 
to a Feynman diagram shown  in Fig.~\ref{t3:pic}.
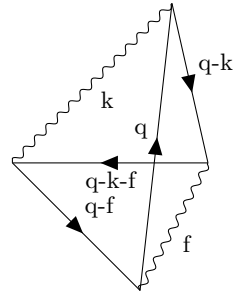
\begin{figure}[ht] 
	\begin{tikzpicture}[baseline=(d)]
	\begin{feynman}[horizontal = (a) to (b)]
	\vertex (d);
	\vertex [right =2.6cm of d] (a) ;
	\vertex [below right =2.4cm of d] (b) ;
	\vertex [above right = 3cm of d] (g) ;
	\diagram*{
		(g) -- [photon, edge label=k] (d) -- [fermion, edge label=q-f] (b),
		(a) -- [fermion, edge label=q-k-f] (d), 
		(a) -- [photon, edge label=f]  (b),
		(g) -- [fermion, edge label=q-k] (a),
		(b) -- [fermion, edge label=q] (g)
	};
	\end{feynman}
	\end{tikzpicture}
	\caption{Toy-model spacetime consisting of 4 fermions and 2 bosons.
		If, for a moment, we consider it as a usual Feynman diagram of 
		quantum field theory and request momentum conservation at each 
		vertex, it will be immediately flattened to a 2d plane.  Redrawn from \cite{Altaisky24psn}}
	\label{t3:pic}
\end{figure}
In the language of spin networks this is a single fermion loop, assembled into strands by  four 4-valent vertices, as  is shown 
in Fig.~\ref{t3f:pic}.
\begin{figure}[ht]
	\centering \includegraphics[width=6cm]{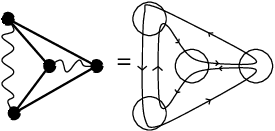}
	\caption{Tetrahedron diagram with 4 spin-$\frac{1}{2}$ fermion lines and two spin-1 boson lines. It can be redrawn as a single 
		loop assembled by four 4-valent vertices  into a strand network. Strands of two antiparallel lines represent bosons. Single lines represent spin-$\frac{1}{2}$ fermions.}
	\label{t3f:pic}
\end{figure} 
The network shown in Fig.~\ref{t3f:pic} is a '2d universe': a boundary of 3d simplex is isomorphic to the sphere $S^2$. 

It would be notoriously difficult to draw a similar combinatorics 
for a 4d universe, but numerical simulations for small random networks 
seem to be similar to the simulation of random surfaces \cite{AJL2005}.

\section{Conclusion}
In this paper we have implemented a construction of a combinatorial 
spacetime solely in terms of matter fields, which meets the original proposal of Roger Penrose \cite{Penrose1971}. The implementation is 
based on the Regge idea that all the curvature of the spacetime manifold 
can be attributed to the vertices of the triangulation graph $\Gamma$ 
\cite{Regge1961}. This assumption is in a striking contrast with the 
usual formulation of loop quantum gravity \cite{RS1995,Vidotto2014,Vaid-book}. By changing the basic object from dual $\tilde{\Gamma}$ to the triangulation graph $\Gamma$ we make the 
model physically tractable, since the vertices of triangulation graph 
have a direct interpretation  of {\em physical events}, with its edges being the matter fields, and the 
whole triangulation graph $\Gamma$ becomes the world history Feynman 
diagram. 

As we can infer from the tetrad construction procedure \eqref{lte},
the proposed model does not give any preferences to the time coordinate. 
This is different from the usual SN formalism, where a network is laid 
in a space-like manifold  and then grows in orthogonal direction to make  spin foam \cite{Baez1998}. The absence of any preferences between directions 
seems advantageous in view of general coordinate invariance \eqref{gct}.

If there are no preferences for the time, there is a natural question: What is the evolution? This question has long been discussed in the context 
of quantum gravity, see, e.g., \cite{Rovelli-book,Vidotto2014} and references therein. The author's point of view is that the role of evolution parameter may be played by the number of degrees of freedom 
in the Universe, appearing by means of the renormalization mechanism. This 
possibility has been already discussed in context of asymptotic
safety \cite{FR2021} and in context of anti-de Sitter spaces 
\cite{AR2022,FR2022}. 

\begin{figure}[ht]
 	\centering \includegraphics[width=5cm]{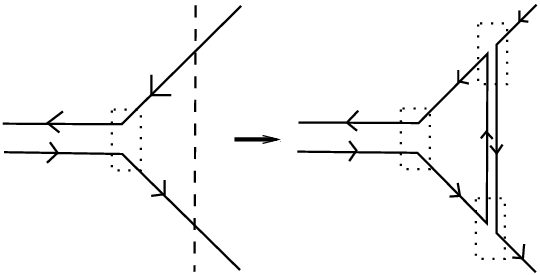}
 	\caption{Renormalization procedure for fusion of two spin-$\frac{1}{2}$ fermions into a spin-1 boson vertex in a 
 	matter spin network}
 	\label{rgp:pic}
 \end{figure}

Renormalization of a fusion vertex in a matter spin network is straightforward and coincides with usual vertex renormalization 
in QED. It is schematically shown in 
 Fig.~\ref{rgp:pic}.
The RG evolution of the matter spin network spacetime will 
be the subject of future studies. It may happen in this approach 
that we live in a universe expanding in all four spacetime directions, 
rather than in FLRW universe expanding in spatial directions. 
The role of cosmological time in such 5d universe may be 
played by RG coordinate \cite{AR2022,FR2022}.
 
\section*{Acknowledgement}
The author is thankful to Dr. K.G.Falls for useful references, to 
Dr. M.Hnatich for useful comments, and  
to Dr. D. Vaid for stimulating discussions.
 %
\end{document}